\documentclass[final,3p,times]{elsarticle}
\usepackage{graphicx}
\usepackage{amssymb}
\pdfoutput=1
\makeatletter
\def\ps@pprintTitle{%
 \let\@oddhead\@empty
 \let\@evenhead\@empty
 \def\@oddfoot{}%
 \let\@evenfoot\@oddfoot}
\makeatother
\biboptions{compress}
\begin{document}
\begin{frontmatter}
\title{Experimental realization of negative refraction and subwavelength imaging for flexural waves in phononic crystal plates}
\author{Hrishikesh Danawe}
\author{Serife Tol\corref{Tol}}
\ead{stol@umich.edu}
\address{Department of Mechanical Engineering, University of Michigan, Ann Arbor, MI USA 48109}
\cortext[Tol]{Corresponding author}

\begin{abstract}
 In this paper, we numerically and experimentally demonstrate negative refraction of flexural waves in phononic crystal (PC) plates which is employed for designing flat elastic lenses. We propose a thickness contrast-based plate design to achieve refractive index equal to -1 at the interface of the PC and host plate. The thickness contrast between the PC and host plate enables matching their wave numbers in the all angle negative refraction (AANR) frequency regime.  The PC-lens design is then numerically and experimentally verified to achieve the image of an omnidirectional subwavelength excitation source. By changing the thickness contrast between the plate and PC, the PC-lens can be tuned for wave focusing and subwavelength imaging at a desired frequency.  
\end{abstract}


\end{frontmatter}

%%\linenumbers

\section{Introduction}
\label{S:1}
Negative refraction resulting from simultaneous negative electrical permittivity and magnetic permeability in left-handed materials was first theoretically predicted by Veselago (1968) \cite{Veselago1968} and seeded the concept of “perfect” lens overcoming the diffraction limit. Based on negative refraction, in his seminal work, Pendry (2000) \cite{Pendry2000} proposed an optical superlens for high resolution imaging. On the other hand, negative refraction phenomenon was also realized via photonic crystals \cite{Notomi2002,Cubukcu2003,Valentine2008,Yao2008,Iyer2002,Smith2004,Joannopoulos1997}. Instead of the double negativity, negative refraction in photonic crystals arises from the intense Bragg scattering described by the band structure and equal frequency contours of the photonic crystals \cite{Li2007}. Similar to perfect lenses with left-handed materials, photonic crystals led to flat lenses enabling subwavelength imaging via all angle negative refraction (AANR), i.e. negative refraction for beams of all incident angles \cite{Parimi2003,Luo2002,Luo2003}. Inspired by the photonic crystals, phononic crystals have taken significant research interest to achieve negative refraction of acoustic or elastic waves. Phononic crystals (PCs) are artificially engineered periodic structures formed by recurring arrangement of inclusions/scatterers in the matrix of base material which results in periodic variation of acoustic/elastic properties of the structure \cite{Kushwaha1993,Pennec2010,Hussein2014}. Analogous to photonic crystals, phononic crystals can exhibit negative refraction with the Poynting vector (group velocity) and wavenumber vector pointing in opposite directions due to bands folding resulting from the periodicity and the negative slope of some optical branches \cite{Pierre1997,Page2016}. AANR is achieved in isotropic or weakly-anisotropic lattices with the dispersion contours being circular and the wavenumbers being same in all directions after a threshold frequency. Hence, flat PC-lenses can be designed for imaging in the AANR frequency regime. Compared to the other wave focusing techniques such as gradient-index PC lenses \cite{Tol2017PC,Tol2019,Jin2019,Danawe2020}, mirrors \cite{Carrara2013,Harne2016,Tol2017Mirrors}, or phased array self-bending elastic/acoustic waves \cite{Zhang2014,Zhu2016,Tol2017SB,Darabi2018,Su2018} which has at least minimum wavelength resolution as their natural limit, PC flat-lenses based on negative refraction have the advantage of acoustic or elastic wave focusing with subwavelength resolution which is highly favorable in medical imaging, non-destructive testing and evaluation, or other applications that require localized wave intensity in an area smaller than a square wavelength.  

Over the last two decades, negative refraction and imaging of acoustic waves have been studied in two-dimensional (2D) PCs \cite{Sukovich2008,Ke2005,Li2006,Lu2007}. For instance, Sukhovich et al. (2008) \cite{Sukovich2008} designed a 2D PC composed of stainless steel rods periodically arranged in water forming a triangular lattice. They matched the wavelengths of sound in water and in PC in the AANR frequency regime and thus dramatically improved the imaging resolution (0.55\(\lambda\)). Also, 2D PCs have been designed for negative refraction of elastic waves in solids. For example, Morvan et al. (2010) \cite{Morvan2010} experimentally demonstrated negative refraction of transverse elastic waves in a 2D PC made up of a square lattice of cylindrical air cavities in an aluminum matrix. Later, Cro\"enne et al. (2011) \cite{Croenne2011} studied negative refraction of longitudinal elastic waves through a prism made up of a 2D solid-solid phononic crystal with triangular lattice arrangement of steel rods in epoxy and they experimentally showed imaging with flat lens immersed in fluid. Besides bulk materials, in elastic wave guides such as plates multiple Lamb wave modes (longitudinal (S), flexural (A) and shear horizontal (SH) modes) can exist at a single frequency and be utilized for negative refraction and imaging. For the first time, Pierre et al. (1997) \cite{Pierre1997} experimentally evidenced the negative refraction of zeroth order flexural Lamb wave mode (A0) in a silicon PC-plate at few MHz. Later, negative refraction of shear horizontal and longitudinal Lamb wave modes were also demonstrated both theoretically and experimentally \cite{Lee2011,Zhu2014}. Employing negative refraction, several researchers investigated imaging via plate-lens designs \cite{Oh2017,Bramhavar2011,Dubois2013}. For instance, Bramhavar et al. (2011) \cite{Bramhavar2011} exploited negative refraction and mode conversion between forward- and backward propagating waves at the homogeneous plate interface with step change in thickness and showed longitudinal Lamb wave focusing at 28 MHz with resolution of 0.98\(\lambda\). Later, Oh et al. (2017) \cite{Oh2017} proposed an isotropic elastic metamaterial for focusing zeroth order longitudinal (S0) Lamb wave mode and experimentally verified subwavelength imaging at 35 kHz with 0.486\(\lambda\) resolution. Furthermore, Dubois et al. (2013) \cite{Dubois2013} utilized the step change in thickness for designing a PC flat-lens to obtain long angle negative refraction (LANR) of zeroth order flexural (A0) Lamb wave mode in the low frequency regime of 5-10 kHz. Unlike the other flat-lens designs utilizing the second (optical) branch of dispersion curves \cite{Sukovich2008,Oh2017}, they only studied imaging with the first (acoustic) branch limiting their design below the first Bragg bandgap. Despite all the efforts in this rapidly progressing field, the negative refraction and imaging for the second dispersion branch of flexural waves has not been well investigated yet. Thus, in the current work, we numerically and experimentally demonstrate the negative refraction of A0 Lamb wave mode for the second dispersion branch (where the direction of the wave vector and group velocity are opposite to each other) and propose a tunable thickness contrast-based PC-lens to achieve refractive index equal to -1 for all angle negative refraction (AANR) based imaging. We also present a thorough discussion on the equal frequency contours and imaging performance of the PC-lens for a subwavelength source.

This paper is organized as follows: In Section \ref{S:2}, unit cell design and dispersion curves of the phononic crystal plate are discussed leading to negative refraction of A0 Lamb wave mode. Section \ref{S:3} provides numerical and experimental verification of negative refraction of A0 mode through a 30\(^{\circ}\)-60\(^{\circ}\)-90\(^{\circ}\) angled prism. In Section \ref{S:4}, a thickness contrast based plate design is proposed and demonstrated to achieve refractive index of -1 by matching the wave numbers in the PC and host plate. Section \ref{S:5} presents numerical and experimental validation of imaging via the proposed design along with the discussions on equal frequency contours. Conclusions are summarized in Section \ref{S:6}.  

\section{Phononic crystal design and negative refraction theory}
\label{S:2}
\subsection{Unit cell design and band structure calculation}

\begin{figure}[h]
\centering\includegraphics[width=0.7\linewidth]{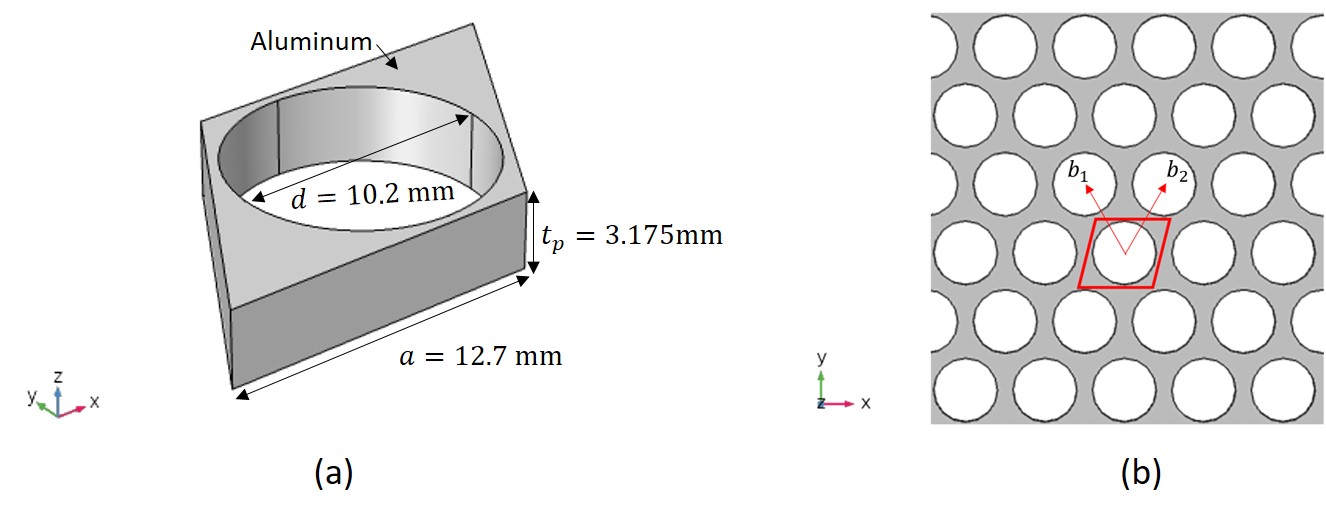}
\caption{(a) Unit cell design: triangular lattice with a through hole. (b) 2D phononic crystal plate and the direct lattice basis vectors \(b_{1}\) and \(b_{2}\) depicted in one unit cell (red box). }\label{fig1} 
\end{figure}

\begin{figure}[h]
\centering\includegraphics[width=0.8\linewidth]{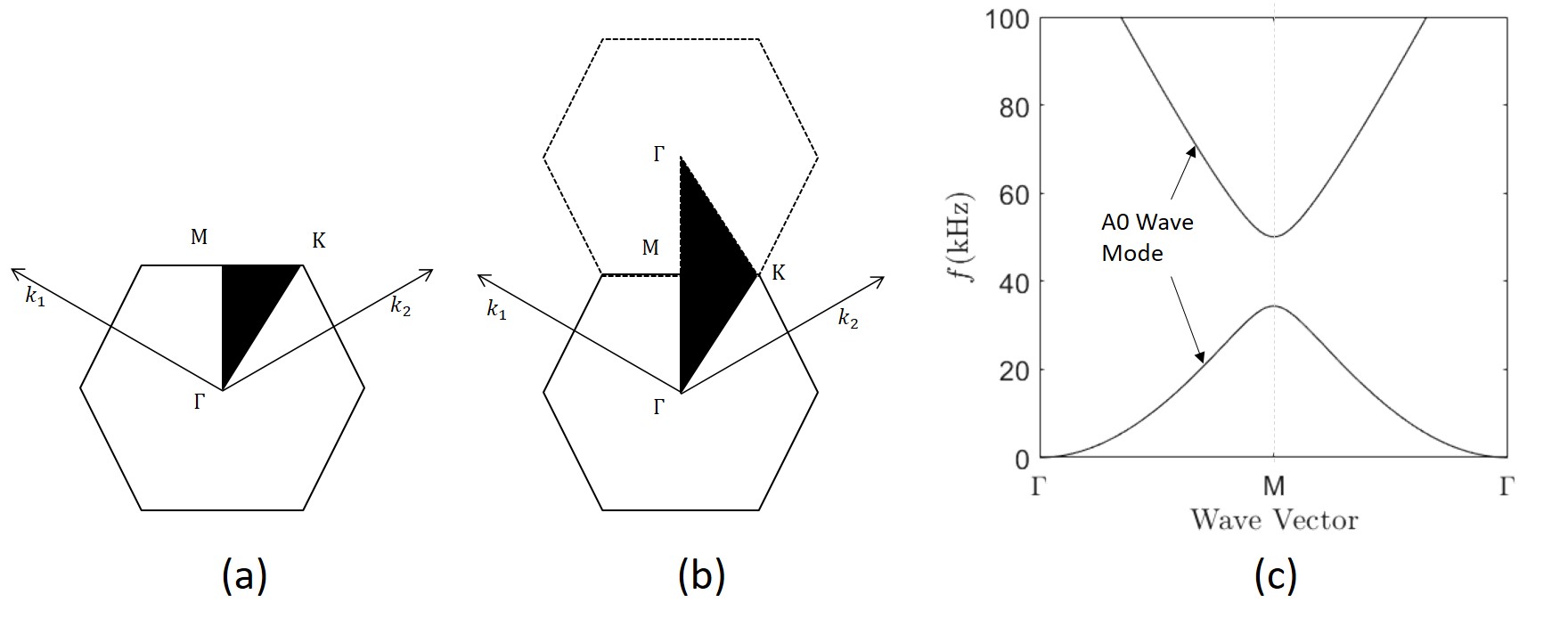}
\caption{(a) The first Brillouin zone of the triangular lattice in reciprocal space where \(k_{1}\) and \(k_{2}\) are reciprocal lattice basis vectors, \(\Gamma\), M and K are lattice symmetry points, and the shaded region is the first irreducible Brillouin zone. (b) Extended Brillouin zone for the triangular lattice. (c) Dispersion curves of A0 Lamb wave mode  plotted along \(\Gamma\)-M-\(\Gamma\) symmetry direction.}\label{fig2}
\end{figure}

The unit cell consists of a central hole drilled in rhombus shaped portion of an aluminum plate as shown in Fig. \ref{fig1}(a). The unit cell size, \(a\), is 12.7 mm and the hole diameter, \(d\), is 10.2 mm. The plate thickness, \(t_{p}\), is chosen as quarter of the unit cell size. The smaller internal angle of rhombus is \(60^{\circ}\), thus resulting in a triangular lattice formation due to tessellation of the unit cell along the basis vectors \(b_{1}\) and \(b_{2}\) of the lattice as illustrated in Fig. \ref{fig1}(b).  
The band structure for the triangular lattice is obtained in COMSOL Multiphysics using solid mechanics module coupled with eigenfrequency study. Floquet-Bloch periodic boundary conditions are applied at unit cell sides. The Floquet periodicity is governed by the equation \cite{Garcia2015}:
\begin{equation}
\label{(eq1)}
\textbf u_{2}=\textbf u_{1}e^{-i \textbf k_{b}.(\textbf r_{2}- \textbf r_{1})}
\end{equation}where \(\textbf u_{1}\), \(\textbf u_{2}\) are displacement vectors and  \(\textbf r_{1}\), \(\textbf r_{2}\) are position vectors on opposite faces along a basis vector, and \(\textbf k_{b}\) is the Bloch wave vector. The dynamics of unit cell is governed by linear elasticity theory and the governing elastodynamic equation for any solid is as follows \cite{Poruchikov1993}:
\begin{equation}
\label{(eq2)}
\rho (\textbf r) \frac{\partial^{2} u_{i}(\textbf r,t)}{\partial t^{2}}=\sum_{ijkl} \frac{\partial }{\partial x_{j}}\left[C_{ijkl}(\textbf r)\frac{\partial u_{l}(\textbf r,t)}{\partial x_{k}}\right]
\end{equation}
where   \(u_{i}(\textbf r,t)\) is a displacement field component as a function of position vector \(\textbf r\) and time \(t\) and \(C_{ijkl}(\textbf r)\) is the elastic tensor and \(\rho (\textbf r)\) is material density both as a function of position vector \(\textbf r\). The finite element discretization of the solid domain can be obtained using Eq. (\ref{(eq2)}) along with boundary conditions of Eq. (\ref{(eq1)}). Consequently, the following eigenvalue problem can be written:\begin{equation}
\label{(eq3)}
\textrm{det}\left(\textrm{\textbf K}(\textbf k_{b})-\omega^2\textrm{\textbf M}(\textbf k_{b})\right)=0
\end{equation}where \(\textrm{\textbf K}\) and \(\textrm{\textbf M}\) are mass and stiffness matrices, respectively, and \(\omega\) is the angular frequency. The mass and the stiffness matrices are both functions of Bloch wave vector \(\textbf k_{b}\). By sweeping the Bloch wave vector in the first Brillouin zone of reciprocal lattice  as shown in Fig. \ref{fig2}(a), one can solve for frequencies resulting in dispersion curves. The dispersion curves for the triangular lattice along \(\Gamma\)-M-\(\Gamma\) crystal symmetry direction of the extended Brillouin zone depicted in Fig. \ref{fig2}(b) are plotted for the lowest anti-symmetric Lamb wave (A0 mode). The dispersion curves are mirrored at the boundary of first and extended Brillouin zone due crystallographic symmetry of triangular lattice. Both branches in Fig. \ref{fig2}(c) correspond to A0 Lamb wave mode and they are separated by the first Bragg bandgap of the PC. The negatively sloped  branch in the first Brillouin zone has a group velocity opposite to the wave vector which results in negative index of refraction as further explained in the next subsection. 

\subsection{Snell's law and negative refraction}

\begin{figure}[h]
\centering\includegraphics[width=0.9\linewidth]{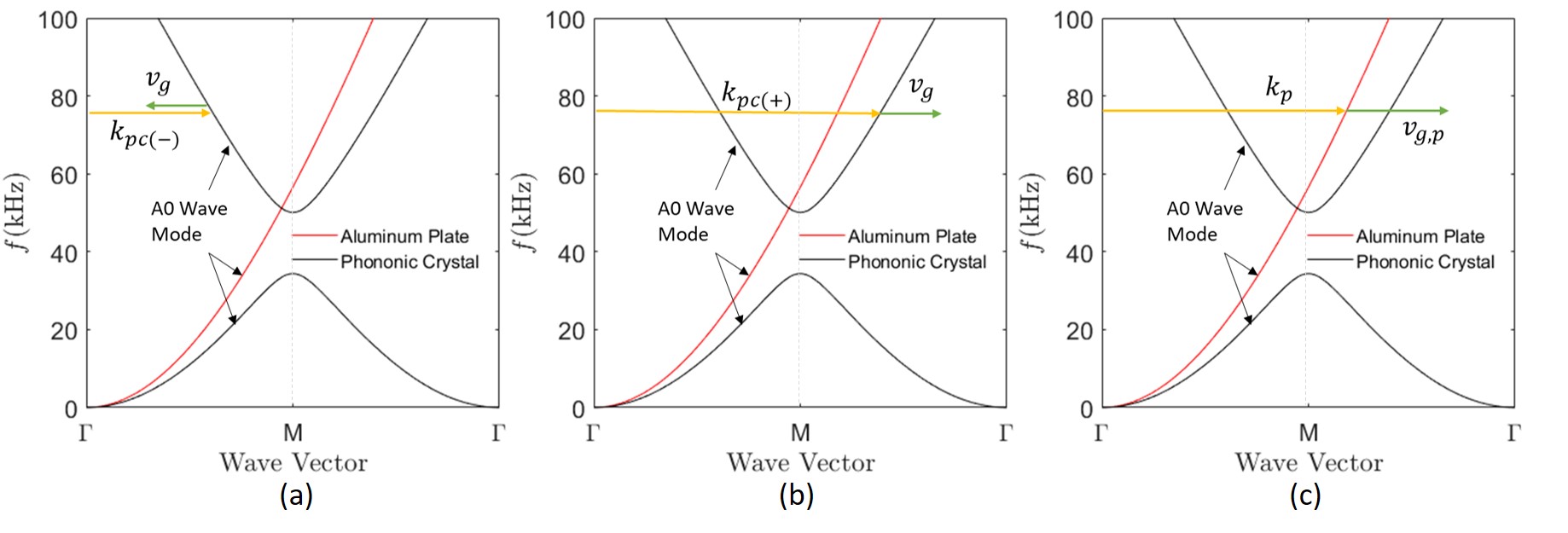}
\caption{Dispersion curves of the aluminum (Al) plate and phononic crystal (PC) with same plate thicknesses. Group velocity vectors and wave vectors are shown (a) for the negatively sloped portion and (b) for the positively sloped portion of the second dispersion branch of the PC, and (c) for the dispersion branch of the aluminum plate.}\label{fig3}
\end{figure}

Figure \ref{fig3} presents the dispersion curves of the homogeneous aluminum plate and the PC plate for the A0 wave mode. The second dispersion branch of the PC has two wave vectors at a given frequency, one for the negatively sloped branch in the first Brillouin zone (\(k_{pc(-)}\) in Fig. \ref{fig3}(a)) and the other for the positively sloped branch in the extended Brillouin zone (\(k_{pc(+)}\) in Fig. \ref{fig3}(b)). The group velocity is defined as the slope of the dispersion branch (i.e \(\partial\omega/\partial k\)). Hence, for the PC, the group velocity \(v_{g}\) is opposite to wave vector \(k_{pc(-)}\) for negatively sloped branch (Fig. \ref{fig3}(a)) and along the wave vector \(k_{pc(+)}\) for the positively sloped branch (Fig. \ref{fig3}(b)). Thus, the waves in the PC with wave vector \(k_{pc(-)}\) has negative refractive index. For the aluminum plate, at the same frequency, the group velocity \(v_{g,p}\) is also along the wave vector \(k_{p}\) as shown in Fig. \ref{fig3}(c).  

\begin{figure}[h]
\centering\includegraphics[width=0.8\linewidth]{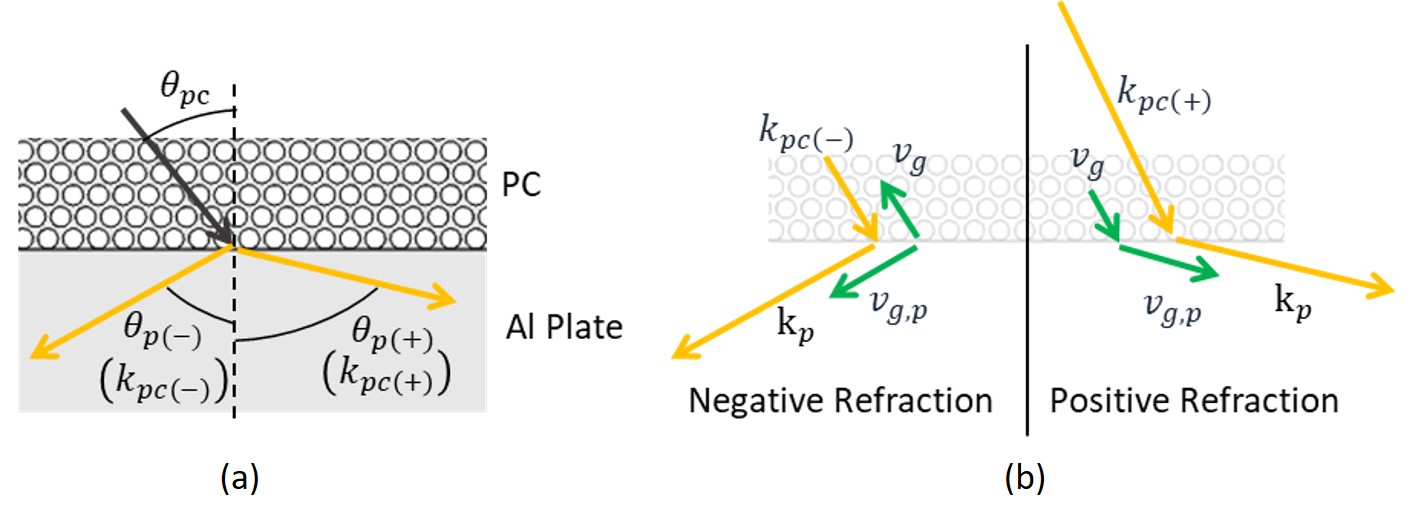}
\caption{(a) Ray diagram depicting wave refraction at the interface of the PC and Al plate. There are two refracted waves in the Al plate corresponding to two wave vectors of the PC (\(k_{pc(-)}\) and \(k_{pc(+)}\)) at a single frequency. (b) The refraction phenomenon explained with wave vectors and group velocity vectors. The negative refraction results when the wave vector points opposite to the group velocity in one of media (in this case, the PC).}\label{fig4}
\end{figure}

The Snell's law of refraction at the interface of two mediums can be expressed as follows:
\begin{equation}
\label{(eq4)}
k_{i}\sin(\theta_{i})=k_{r}\sin(\theta_{r})
\end{equation}
where \(k_{i}\) and \(k_{r}\) are wave vectors of the incident and refracted waves and \(\theta_{i}\) and \(\theta_{r}\) are angles of incidence and refraction, respectively.  Considering an interface between the PC and aluminum plate as shown in Fig. \ref{fig4}(a), the incident wave impinging on the PC side results in two refracted waves on the homogeneous plate side. One is positively refracted wave corresponding to \(k_{pc(+)}\) wave vector and the other is negatively refracted wave corresponding to \(k_{pc(-)}\) wave vector in the PC. The wave vectors and group velocities for both types of refraction are depicted in Fig. \ref{fig4}(b). The negative refraction phenomenon occurs for the wave whose wave vector is opposite to its group velocity. 

\section{Simulations and experiments: Negative refraction of A0 Lamb wave mode}
\label{S:3}
\subsection{Prism design and numerical simulations for negative refraction}

\begin{figure}[h]
\centering\includegraphics[width=0.7\linewidth]{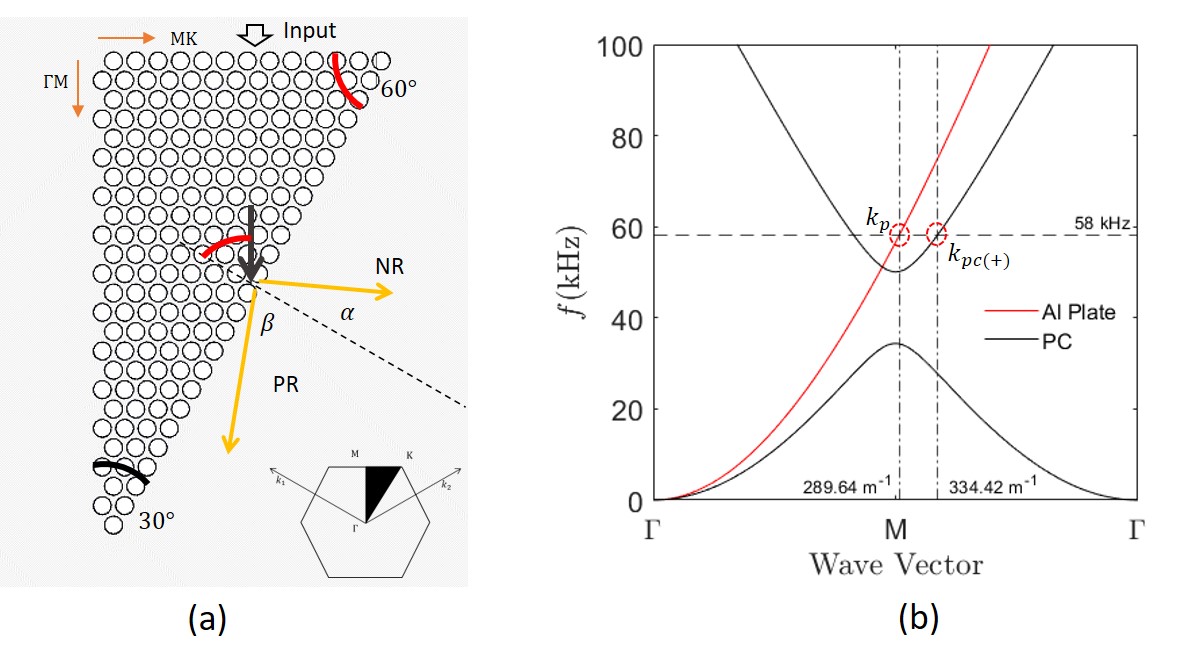}
\caption{(a) Prism design illustrating refraction phenomenon at the longest side of the triangle with incident angle of 60\(^{\circ}\) (red curve), \(\alpha\) and \(\beta\) are refraction angles for negative and positive refraction, respectively,  and \(\Gamma\)M and MK are crystal symmetry directions as shown in the Brillouin zone (bottom right). (b) Dispersion curves showing magnitudes of wave vectors \(k_{pc(+)}\) and \(k_{p}\) at 58 kHz for the critical angle of positive refraction (\(\beta = 90^{\circ}\)).}\label{fig5}
\end{figure}

In order to demonstrate the negative refraction of flexural waves travelling in \(\Gamma\)M crystal symmetry direction, a 30\(^{\circ}\)-60\(^{\circ}\)-90\(^{\circ}\) angled triangular prism is designed such that the sides of the triangle are aligned with the crystal symmetry directions of the first irreducible Brillouin zone. The prism has 25 unit cells along its longest edge as shown in Fig. \ref{fig5}(a). The \(\Gamma\)M symmetry direction is perpendicular to the shortest edge of the triangle; hence, the waves incident on the shortest edge travels along \(\Gamma\)M direction. The refraction phenomenon happens at the longest edge of the prism and the angle of incidence is fixed as 60\(^{\circ}\)  due to prism angles. Thus, the Snell's law for refraction through the prism becomes:
\begin{equation}
\label{(eq5)}
k_{pc}\sin(60^{\circ})=k_{p}\sin(\theta_{p})
\end{equation}
where \(\theta_{p}\) is the angle of refraction on the host plate  and \(k_{pc}\) and \(k_{p}\) are wave vectors of waves in the PC and Al host plate, respectively. Depending on the combination of the wave vectors, there will be positively and/or negatively refracted waves with angles of refraction \(\beta\) and \(\alpha\), respectively. From Fig. \ref{fig5}(b), it can be observed that magnitudes of wave vectors \(k_{pc(+)}\) and \(k_{p}\) increase with frequency while the magnitude of \(k_{pc(-)}\) decreases with frequency. Thus, there exist a critical frequency above which there is no positively refracted propagating wave possible according to Eq. (\ref{(eq5)}). The critical frequency can be determined by setting \(\beta\) to 90\(^{\circ}\).  The combination of wave vectors \(k_{pc(+)}\) and \(k_{p}\) where the sine of  \(\beta\) becomes 1 is obtained at 58 kHz as shown in Fig. \ref{fig5}(b). Whereas, for wave vectors \(k_{pc(-)}\) and \(k_{p}\), the sine of \(\alpha\) is still less than 1 and decreases as the frequency is increased. Hence, at frequencies above 58 kHz waves are only negatively refracted. 

\begin{figure}[h]
\centering\includegraphics[width=0.8\linewidth]{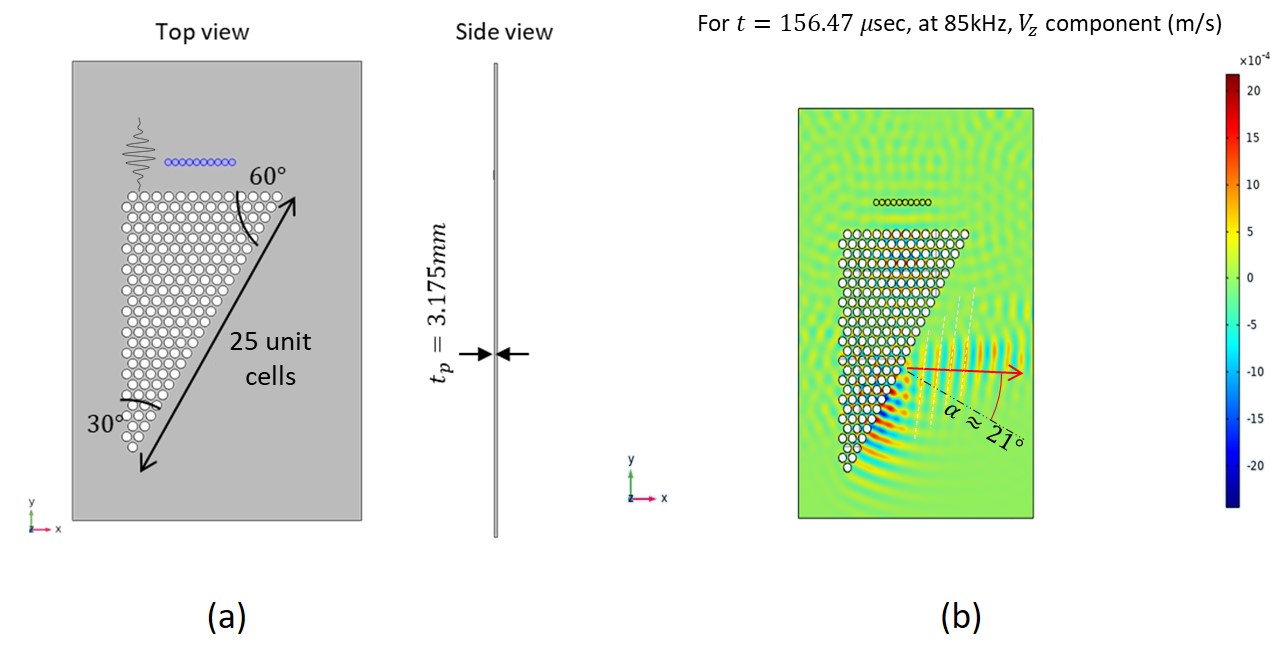}
\caption{(a) Numerical model with array of holes arranged in triangular lattice geometry forming a 30\(^{\circ}\)-60\(^{\circ}\)-90\(^{\circ}\) prism, 10 circular piezoelectric actuators are shown in blue color. (b) Instantaneous velocity field of the out-of-plane component \(V_{z}\) at time \(t=\)156.47 \(\mu\)sec, \(\alpha\) is the angle of negative refraction.}\label{fig6}
\end{figure}

In order to verify the negative refraction phenomenon for A0 Lamb wave mode through the triangular prism design, we performed time-domain numerical simulations in COMSOL Multiphysics. The simulation model shown in Fig. \ref{fig6}(a) was constructed on an Al plate  with a thickness of 3.175 mm (same as the unit cell thickness in Fig. \ref{fig1}(a)) including 25 unit cells on the longest side of the prism. The flexural wave (A0 mode) was excited using an array of 10 circular thickness mode piezoelectric actuators. The piezoelectric actuators were excited using a 7-cycle sine burst signal at a center frequency of 85 kHz which is above the critical frequency. The mesh size was chosen to be 20 times smaller than the wavelength of the excited wave and the time stepping was set to satisfy Courant–Friedrichs–Lewy (CFL) number of 0.2. In order to reduce reflections from plate boundaries, low-reflecting boundary condition was applied at all four sides of the plate. The simulation was run for 200 \(\mu\)sec and the refracted waveform was captured at 156.47 \(\mu\)sec as shown in Fig. \ref{fig6}(b). The direction of wave vector can be determined from the instantaneous out-of-plane velocity field by identifying the wave fronts of the refracted waves and drawing normal to them. Accordingly, the angle of negative refraction measured between the normal to the longest side of the prism and the wave vector is \(\alpha\approx21^{\circ}\). At the excitation frequency of 85 kHz, the magnitude of wave vectors \(k_{pc(-)}\) and \(k_{p}\) from the dispersion curves are obtained as 145.67 m\(^{-1}\) and 359.9 m\(^{-1}\), respectively, resulting in negative angle of refraction as \(\alpha=20.52^{\circ}\) calculated by using Eq. (\ref{(eq5)}). Hence, the refraction angle from simulation results match very well with the theoretical value obtained from dispersion curves. Note that, for the other wave with wave vector \(k_{pc(+)}\), the sine of positive refraction angle, \(\beta\), turns out to be greater that 1 meaning that there is no positively refracted propagating waves. In this case, the refracted wave is an evanescent wave whose amplitude decays away from the interface and into the host plate. We can observe the positively refracted evanescent wave along the interface of the prism and host plate as presented in Fig. \ref{fig6}(b). 

\subsection{Experimental verification of negative refraction through the prism}

\begin{figure}[h]
\centering\includegraphics[width=0.9\linewidth]{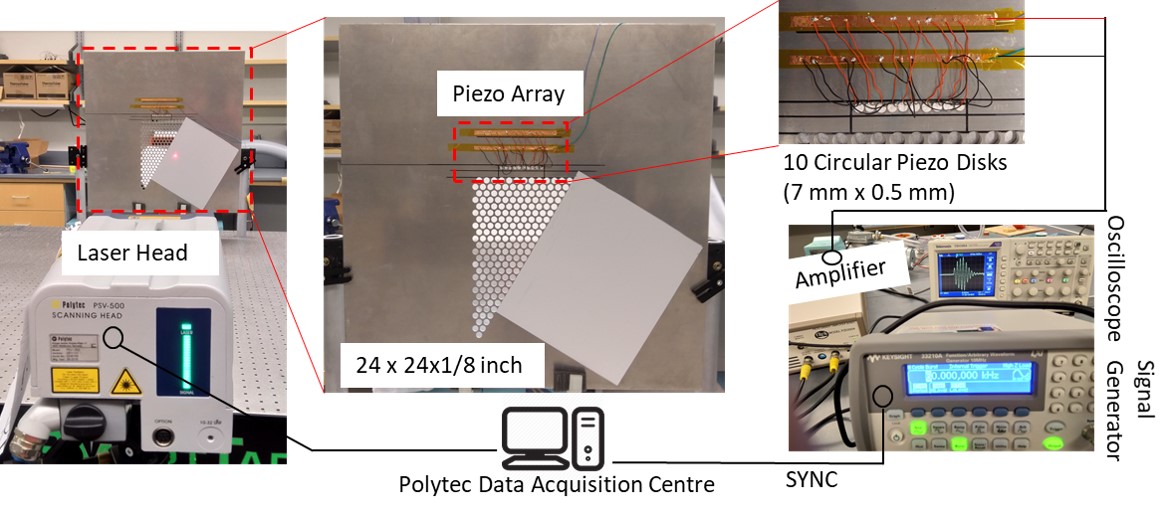}
\caption{Experimental setup for the validation of negative refraction of flexural wave mode through triangular prism embedded in an aluminum plate.}\label{fig7}
\end{figure}

Next, we validate the negative refraction of A0 Lamb mode through the prism design. The triangular prism design made out of PC was embedded in a 609.6 mm x 609.6 mm x 3.175 mm aluminum plate. The plate was chosen big enough to avoid any boundary reflection interfering in the measurement region of the plate. Figure \ref{fig7} shows the overall experimental setup to measure the out-of-plane velocity over the plate surface using PSV 500 Scanning Laser Vibrometer by Polytec Inc. A reflective tape was stuck over the plate surface in the measurement region for accurate laser readings. For flexural wave excitation, we used an array of 10 circular (7 mm x 0.5 mm), thickness-mode vibration piezoelectric discs from Steminc Inc. The piezoelectric disc were bonded over the plate surface using epoxy by employing vacuum bonding technique to ensure perfect contact. The piezoelectric actuators were excited using a Trek power amplifier connected to a Keysight signal generator providing 4-cycle sine burst signal. The signal generator and the scanning laser head were in sync through Polytec data acquisition system. The data acquisition sampling frequency was set at 6.25 MHz taking total of 5000 samples over a duration of 800 \(\mu\)sec. After scanning all grid points, wave fronts of out-of-plane vibration over the scanned region of the plate  were generated in the Polytec post-processing software.    

\begin{figure}[h]
\centering\includegraphics[width=0.9\linewidth]{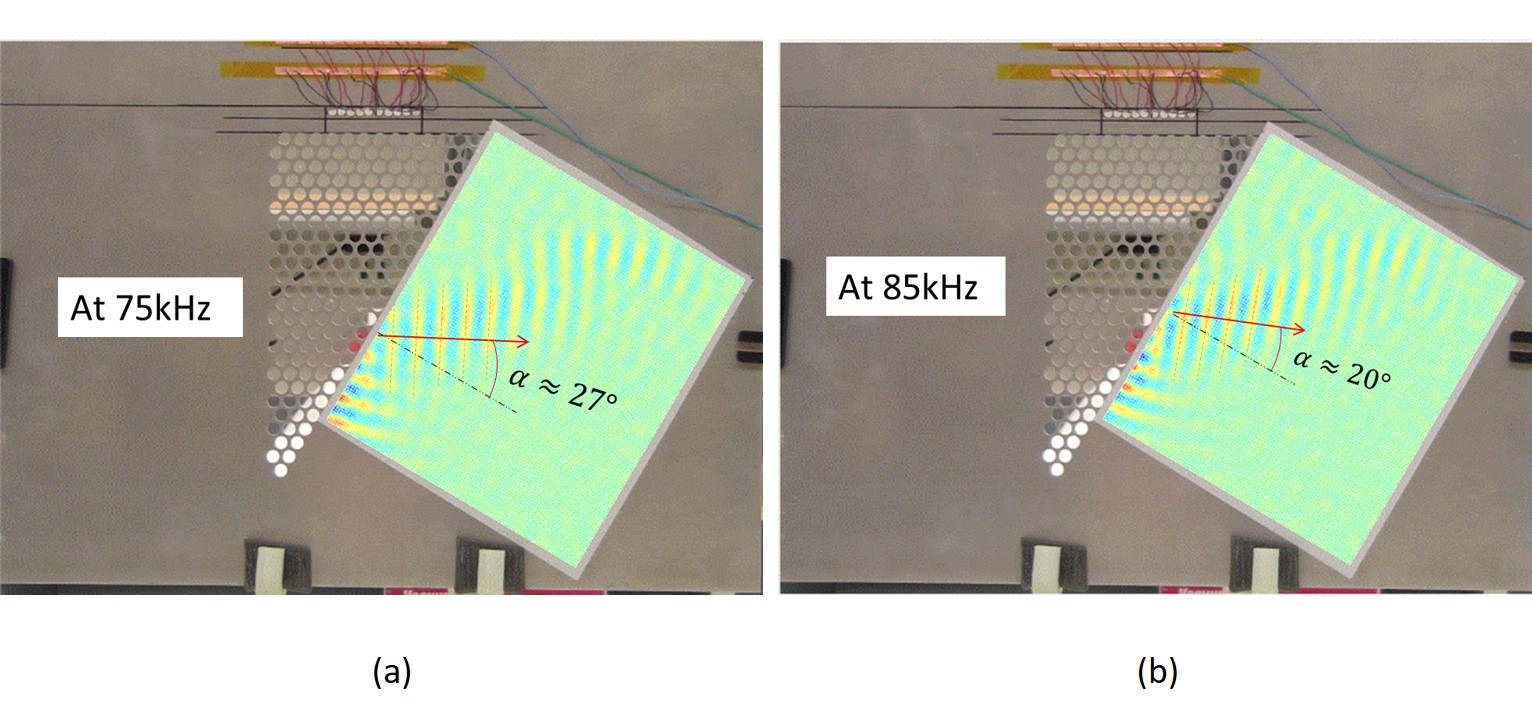}
\caption{Instantaneous out-of-plane velocity field measured over the plate surface at (a) 75 kHz and (b) 85 kHz. The red arrow shows the direction of wave vector and \(\alpha\) is the angle of negative refraction.}\label{fig8}
\end{figure}

The experiments were run at multiple frequencies above 58 kHz for which only negative refraction occurs. The instantaneous out-of-plane velocity field measured over the plate surface at two different frequencies are shown in Fig. \ref{fig8}. The angle of negative refraction, \(\alpha\), is determined by identifying the wave fronts of the refracted waves, following the procedure as explained in Section \ref{S:3}.1. The angle of negative refraction can also be determined from dispersion curves by evaluating the magnitudes of \(k_{pc(-)}\) and \(k_{p}\) at the excitation frequency (similar to Fig. \ref{fig5}(b)). The theoretical negative angles of refraction at 75 kHz and 85 kHz are \(\alpha=27.06^{\circ}\) and \(\alpha=20.52^{\circ}\), respectively, as obtained from dispersion curves using Eq. (\ref{(eq5)}). Experimental values of negative refraction angles measured at 75 kHz (\(\alpha\approx27^{\circ}\)) and 85 kHz (\(\alpha\approx20^{\circ}\)) are in excellent agreement with the theoretical predictions.
 
\section{Thickness contrast-based PC-lens design for imaging}
\label{S:4}

\begin{figure}[h]
\centering\includegraphics[width=0.8\linewidth]{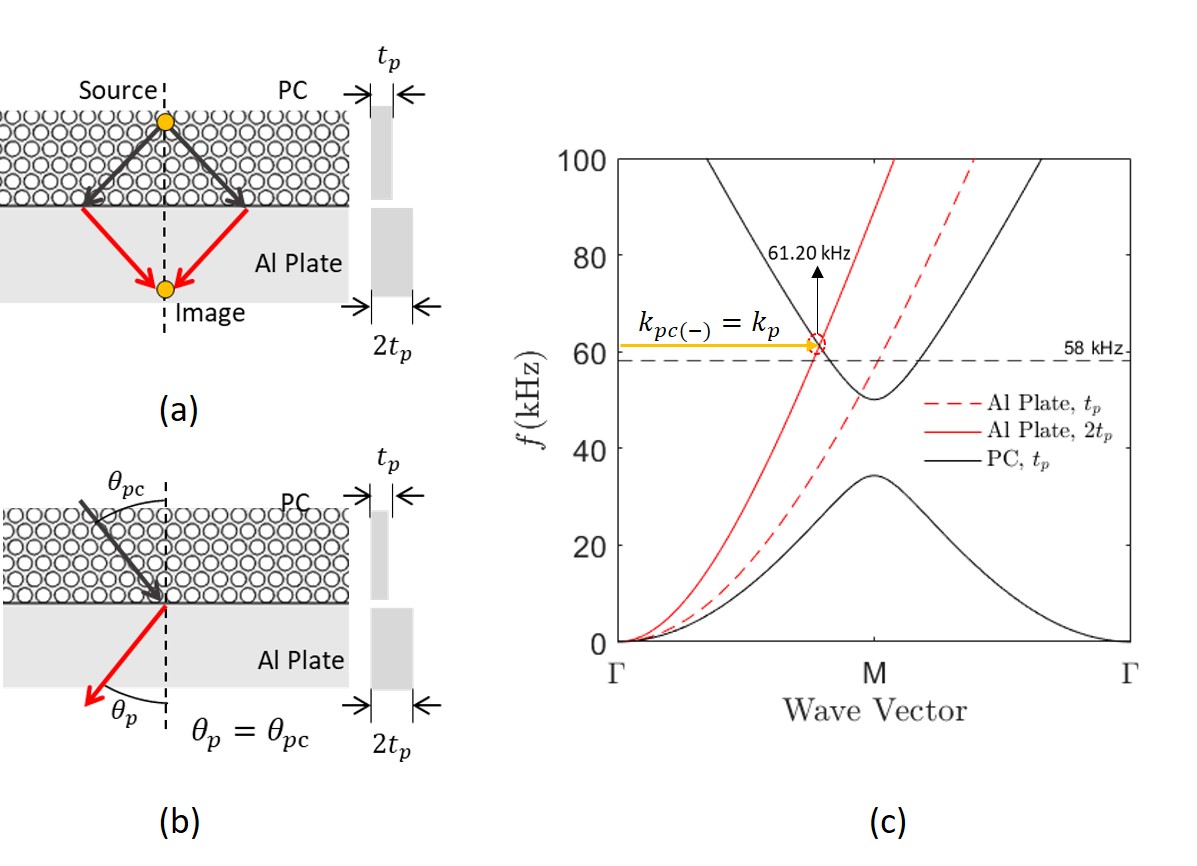}
\caption{(a) Schematic demonstration of imaging using thickness contrast design at 61.20 kHz (for which the refractive index is -1). (b) Negative refraction at PC-Al plate interface at 61.20 kHz for which the wave vectors \(k_{pc(-)}\) and \(k_{p}\) are equal. (c) Dispersion curves for homogeneous Al plates with thickness \(t_{p}\) and \(2t_{p}\) plotted along with the dispersion curves of PC with the plate thickness of \(t_{p}\).}\label{fig9}
\end{figure}

The negative refraction phenomenon at the interface of the aluminum plate and PC can be utilized to design flat elastic lenses for wave focusing or imaging applications. In order to obtain a sharp image, all waves originating from the excitation source should focus at a single point after passing through the lens as depicted in Fig. \ref{fig9}(a). This can be achieved if the refractive index at the interface of Al plate and PC is equal to -1 for all angles of incidence which is called all angle negative refraction (AANR). The refractive index of -1 results in equal angles of incident and refraction on the same side of the normal at the interface as shown in Fig. \ref{fig9}(b). From Eq. (\ref{(eq4)}), the refractive index of -1 is achieved when the wave vectors in the Al plate and PC are of equal magnitude in the first Brillouin zone for the second dispersion branch of PC (i.e.  \(k_{pc(-)}=k_{p}\)). The wave vector is equal in all directions for homogeneous aluminum plate. Hence, AANR is possible if the wave vector is equal in all directions for the PC as well, such that, the equal frequency contours (EFC) are circular in the first Brillouin zone. The EFC for the triangular lattice PC (as shown in Fig. \ref{fig1}) shows nearly circular pattern for the second dispersion branch as presented in Section \ref{S:5}.3.

The two wave vectors \(k_{pc(-)}\) and \(k_{p}\) can be tuned to be equal at a desired frequency well above the bandgap by introducing a thickness contrast between the Al plate and PC plate as illustrated in Fig. \ref{fig9}. In this study, we chose the plate thickness of Al plate two times that of the PC which shifts the dispersion curves of Al plate from dotted red curve to solid red curve in Fig. \ref{fig9}(c). In this design, the wave vectors are equal at 61.20 kHz above the critical frequency of 58 kHz and thus the index of refraction is -1 at the interface of Al plate and PC for the flexural waves propagating at that frequency. Alternatively, the plate thickness of PC can be changed to \(t_{p}/2\) keeping the Al plate thickness equal to \(t_{p}\). This would shift the dispersion curves of PC towards lower frequency region making it possible to match the wave vectors at lower frequencies. Overall, with the thickness contrast design, the refractive index of -1 can be achieved at a desired frequency in a broad frequency region. Note that, in general the phononic crystal is anisotropic and the wave vector of the isotropic Al plate can be matched with that of PC only along one direction at a time. This will be further explained in Section \ref{S:5} with equal frequency contours and its implications on imaging performance will be discussed through numerical and experimental results.   
\section{Simulations and experiments: Imaging via negative refraction based PC-lens}
\label{S:5}
\subsection{Numerical simulations for imaging with thickness contrast PC-lens design }

\begin{figure}[h]
\centering\includegraphics[width=1\linewidth]{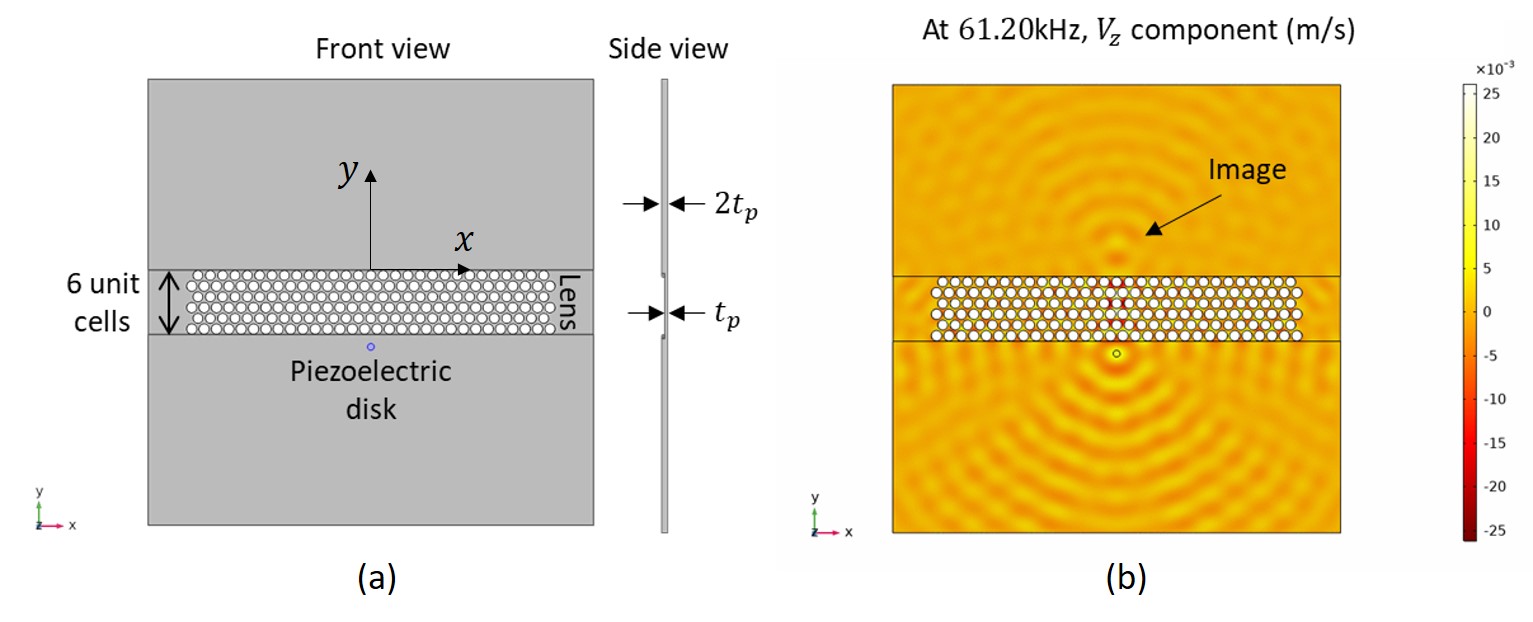}
\caption{(a) Aluminum plate model embedded with 6 unit cell wide flat PC-lens and a piezoelectric actuator (blue) acting as a omnidirecttional source. (b) Instantaneous out-of-plane velocity field \(V_{z}\) for continuous sine wave excitation at 61.20 kHz verifying the image of excitation source (piezoelectric disk) formed on the other side of the lens.}\label{fig10}
\end{figure}

A flat lens made up of PC was embedded into an aluminum plate of size 457.20 mm x 457.20 mm x 6.35 mm as shown in Fig. \ref{fig10}(a).  The thickness contrast was introduced by reducing the thickness of the lens region of the plate up to a thickness of \(t_{p}=3.175\) mm. The triangular lattice formation was created according to the PC design with through holes (Fig. \ref{fig1}). The lens was designed with 6 unit cells in y-direction. A  thin circular piezoelectric disk of size 7 mm x 0.5 mm was attached to the host plate surface on one side of the lens. The time-domain numerical simulations were performed with simulation settings similar to the case of negative refraction explained in Section \ref{S:3}. The piezoelectric actuator was excited at the design frequency of 61.20 kHz. The instantaneous velocity presented in Fig. \ref{fig10}(b) verifies the refracted flexural waves forming an image of the subwavelength excitation source (piezoelectric disk, \(d_{p}\approx 0.25\lambda\)) under continuous wave excitation.        

\begin{figure}[h]
\centering\includegraphics[width=1\linewidth]{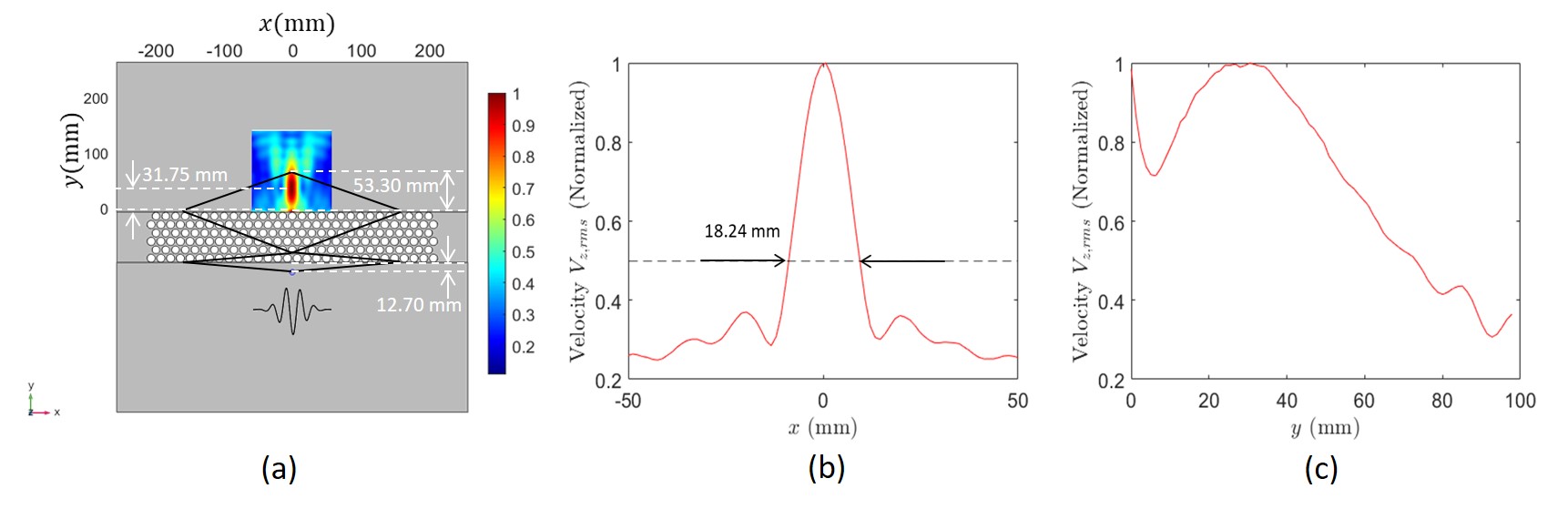}
\caption{ (a) Normalized RMS velocity (out-of-plane) showing the image intensity of the excitation source (piezoelectric disk) excited using 4-cycle sine burst excitation at 61.20 kHz. The line plots along the lines parallel to (b) x and (c) y-axes passing through the maximum intensity point.}\label{fig11}
\end{figure}

The normalized RMS velocity plots for a 4-cycle sine burst excitation are shown in Fig. \ref{fig11}. The maximum intensity in simulations is obtained at 31.75 mm from the upper edge of the lens as compared to the theoretical image location of 53.30 mm calculated from the ray trajectories as illustrated in Fig. \ref{fig11}(a). This mismatch in the image location as well as the longer image in y-direction is further investigated and discussed in Section \ref{S:5}.3 along with the experimental results. The line plots along the lines parallel to x and y-axes passing through the maximum intensity point are plotted in Fig. \ref{fig11}(b) and (c), respectively. The full width at half maximum (FWHM) is measured as 18.24 mm. It is 0.65\(\lambda\) because the wavelength of the 61.20 kHz A0 wave mode of 6.35 mm thick aluminum plate is \(\lambda=27.97\) mm. The maximum RMS velocity obtained in the image region with PC-lens is around 3 times higher than the RMS velocity at the same location without the PC-lens (i.e. baseline measurement). Thus, the PC-lens can focus the flexural wave energy with nearly subwavelength resolution and high intensity (9 times higher than the baseline case). We further validate these results through experiments in the next subsection.

\subsection{Experimental verification of imaging with thickness contrast PC-lens design }

\begin{figure}[h]
\centering\includegraphics[width=0.9\linewidth]{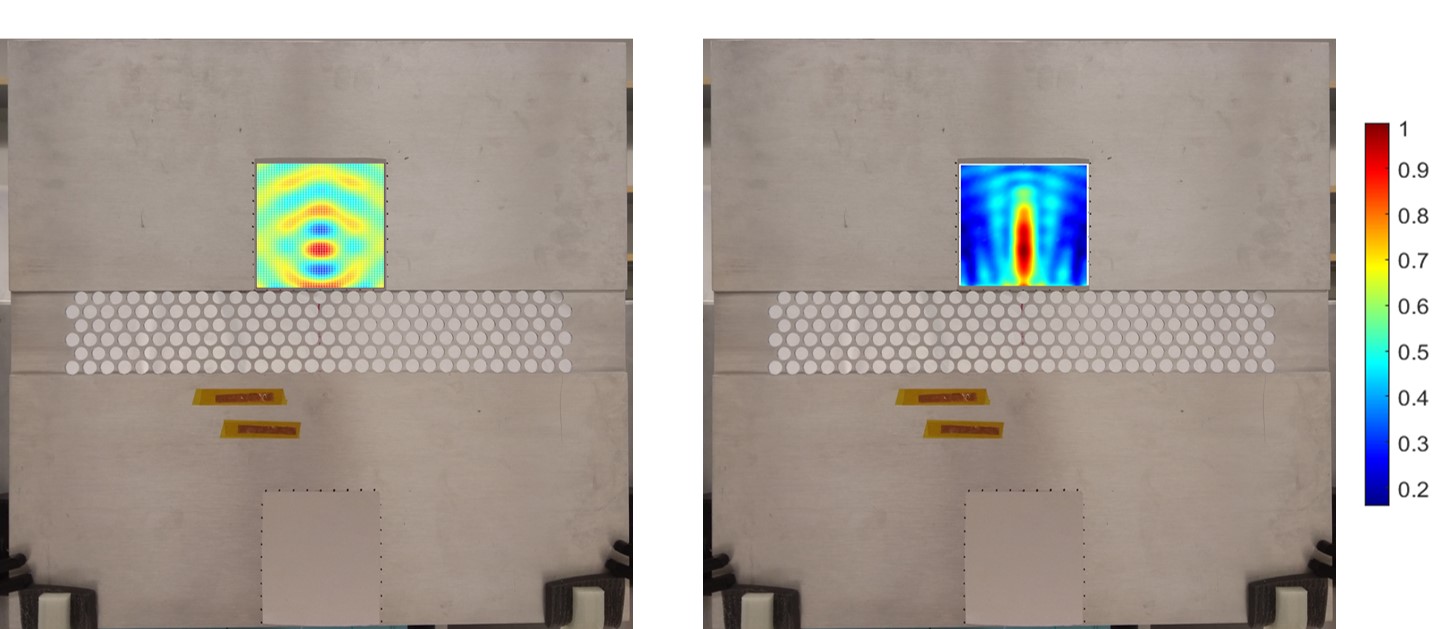}
\caption{(a) Instantaneous out-of-plane velocity field over the plate surface measured at time \(t=\)108 \( \mu\)sec under 4-cycle sine burst excitation at 61.20 kHz.  (b) Normalized RMS velocity field obtained by post-processing the out-of-plane velocity measurements with the scanning LDV.}\label{fig12}
\end{figure}

The aluminum plate manufactured for experiments has the same specifications as the model created for numerical simulations. Also, the similar experimental setup (Fig. \ref{fig7}) was used for imaging measurements in the aluminum plate embedded  with PC-lens. For this experiment, only one circular piezoelectric disk was used as an omnidirectional source. The piezoelectric disk was excited using 4-cycle sine burst at design frequency of 61.20 kHz. Figure \ref{fig12}(a) shows the instantaneous out-of-plane velocity field measured at time \(t=\)108 \( \mu\)sec via the post-processing software by Polytec. Then, the normalized RMS velocity field  as shown in Fig. \ref{fig12}(b) was calculated in MATLAB. The location of the maximum intensity point is determined to be at 31.70 mm away from the upper edge of the lens which matches very well with the numerical results estimated as 31.75 mm.         

\subsection{Comparison of numerical and experimental results and discussions}

\begin{figure}[h]
\centering\includegraphics[width=1\linewidth]{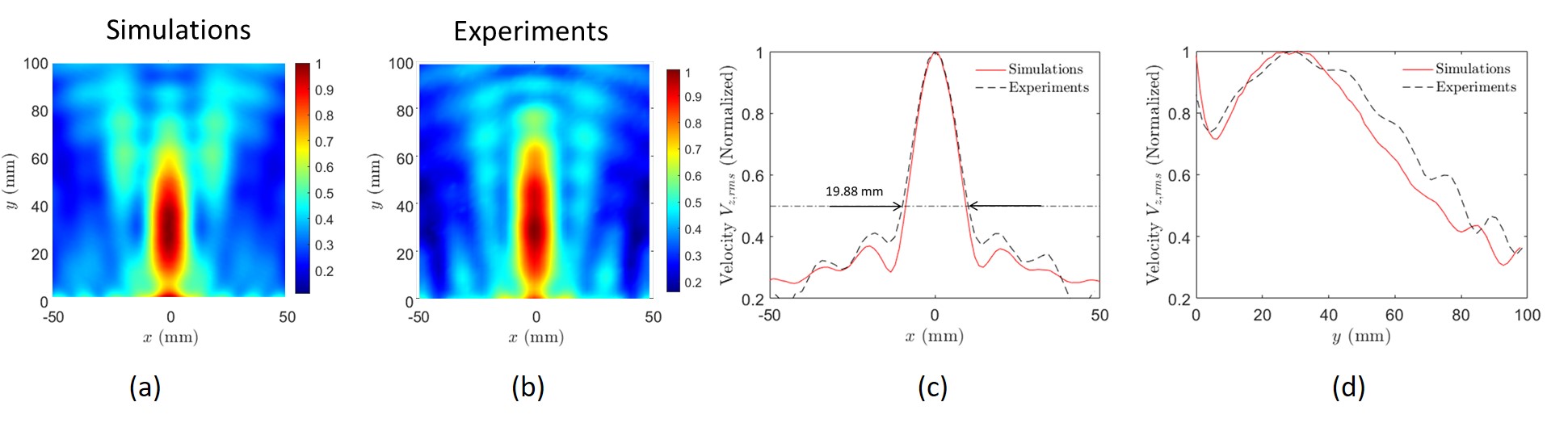}
\caption{The normalized RMS velocity field obtained under 4-cycle sine burst excitation at 61.20 kHz in (a) simulations and (b) experiments. The comparison of line plots from simulations and experiments along the lines parallel to (b) x and (c) y-axes passing through the maximum intensity point.}\label{fig13}
\end{figure}

\begin{figure}[h]
\centering\includegraphics[width=0.9\linewidth]{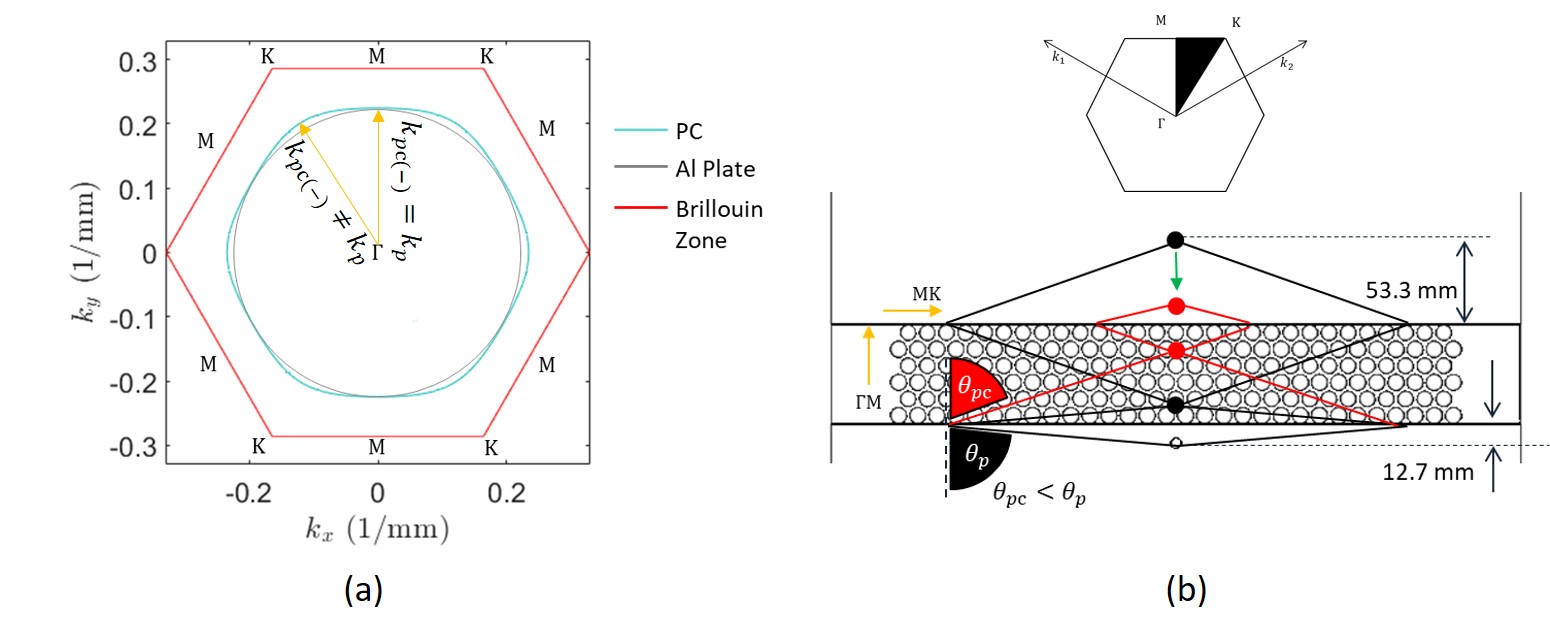}
\caption{(a) Equal frequency contours for A0 wave mode of the Al plate and PC at 61.20 kHz along with the first Brillouin zone of the PC lattice, (the wave vectors \(k_{pc(-)}\) and \(k_{p}\) are equal only along \(\Gamma\)M symmetry direction). (b) The shifting of image due to mismatch of wave vectors along the directions other than \(\Gamma\)M symmetry direction.}\label{fig14}
\end{figure}

Normalized RMS velocity fields obtained in simulations and experiments are compared in Fig. \ref{fig13}. The pattern of the wavefield and the shape of the image are in excellent agreement in simulations and experiments. FWHM measured in experiments equals to 0.71\(\lambda\) that favorably agrees well with numerically estimated FWHM, 0.67\(\lambda\) . Measured FWHM indicates that the proposed PC-lens operates close to the diffraction limit (0.5\(\lambda\)). In addition, the location of the maximum intensity point is experimentally measured at 31.70 mm which matches well with the numerical value predicted as 31.75 mm. Moreover, the deviation of image towards the upper edge of the lens away from the theoretical predictions from ray trajectory (Fig. \ref{fig11}(a)) can be explained with crystal anisotropy. The wave speed in phononic crystal is, in general, not equal in all directions due to crystal anisotropy. The equal frequency contours (EFC), obtained from unit cell simulations, for the A0 wave mode of the Al plate and PC are depicted in Fig. \ref{fig14}(a) along with the first Brillouin zone of the PC lattice. As observed in the EFC plots, the contours are perfect circles for the Al plate suggesting its isotropic nature, meaning that the wave vector (and thus the wave speed) is same in all directions, whereas the contour for PC is not a perfect circle. Hence, even though the PC-lens is ideally designed to have equal wave vectors \(k_{pc(-)}\) and \(k_{p}\) along \(\Gamma\)M symmetry direction, there is no guarantee that the wave vectors are equal in other directions. For the current design, the wave vector \(k_{pc(-)}\) in the directions other than \(\Gamma\)M is bigger than the wave vector \(k_{p}\) resulting in \(\theta_{pc(-)}<\theta_{p}\) according to the Snell's law (Eq. (\ref{(eq4)})). Thus, for the present design, the ray trajectories show that the image shifts closer to the upper edge of the lens as shown in Fig. \ref{fig14}(b). The elongated shape of image along y-direction can also be explained due to wave vector mismatch. As a result of unequal incident and refracted angles, the waves incident at different angles form image at different locations resulting in elongated intensity map of the image. The anisotropy of phononic crystals can be reduced with careful design and optimization of the unit cell geometry. The fact that FWHM=19.88 mm by employing the negative refraction of flexural waves in the PC-plate (wavelenghth, \(\lambda=\) 27.97mm) demonstrates the experimental evidence of enhanced image resolution that is not possible to achieve through conventional wave focusing concepts like GRIN-PC lenses.  Although, this does not break the diffraction limit of \(\lambda/2\) as previously observed by Sukhovich et al. (2008) \cite{Sukovich2008} for ultrasound waves which inspired the lattice design in the present work, it could be possible to break the diffraction limit using other lattice structures through design optimization which will be a future study. Yet, the primary goal of the present work is to provide experimental evidence of negative refraction and imaging of flexural waves (optical branch) for the first time and to propose a thickness contrast-based PC-lens design for wave focusing with subwavelength resolution that can be tunable over a broad frequency range.

\section{Conclusions}
\label{S:6}
We demonstrated negative refraction of A0 mode using a phononic crystal plate with the triangular lattice geometry. Experimental results of negative refraction are in excellent agreement with theoretical and numerical results. Then, we exploited the negative refraction phenomenon to design a high resolution flat lens. We proposed a thickness contrast design to achieve refractive index of -1 by matching wave vectors of the PC and homogeneous aluminum plate at the design frequency. The thickness contrast-based PC-lens design can be easily tuned at a desired design frequency in a broad frequency region. We validated imaging of a subwavelength source via negative refraction with the proposed PC-lens design through numerical simulations and experimental measurements with excellent agreement. The PC-lens can successfully focus the flexural waves with subwavelength resolution and increase the intensity of wave energy by 9 times. We further investigated the effect of crystal anisotropy on imaging performance along with the discussion on the preexisting literature. The proposed frequency tunable design can benefit  many applications such as ultrasonic inspection, tetherless energy transfer, and energy harvesting, where the localization of wave energy in a small spot is desirable.    

\section*{Acknowledgment}
This work was supported in part by the National Science Foundation [grant number CMMI-1914583].



%% References



%% New version of the num-names style
\bibliographystyle{elsarticle-num-names}
\bibliography{Bibliography.bib}

\begin{thebibliography}{43}
\expandafter\ifx\csname natexlab\endcsname\relax\def\natexlab#1{#1}\fi
\providecommand{\url}[1]{\texttt{#1}}
\providecommand{\href}[2]{#2}
\providecommand{\path}[1]{#1}
\providecommand{\DOIprefix}{doi:}
\providecommand{\ArXivprefix}{arXiv:}
\providecommand{\URLprefix}{URL: }
\providecommand{\Pubmedprefix}{pmid:}
\providecommand{\doi}[1]{\href{http://dx.doi.org/#1}{\path{#1}}}
\providecommand{\Pubmed}[1]{\href{pmid:#1}{\path{#1}}}
\providecommand{\bibinfo}[2]{#2}
\ifx\xfnm\relax \def\xfnm[#1]{\unskip,\space#1}\fi
%Type = Article
\bibitem[{Veselago(1968)}]{Veselago1968}
\bibinfo{author}{V.~G. Veselago},
\newblock \bibinfo{title}{{The} {electrodynamics} {of} {substances} {with}
  {simultaneously} {negative} {values} {of} {\(\epsilon\)} {and} {\(\mu\)}},
\newblock \bibinfo{journal}{Sov. Phys. Uspekhi} \bibinfo{volume}{10}
  (\bibinfo{year}{1968}) \bibinfo{pages}{509--514}.
  \DOIprefix\doi{https://doi.org/10.1070/pu1968v010n04abeh003699}.
%Type = Article
\bibitem[{Pendry(2000)}]{Pendry2000}
\bibinfo{author}{J.~B. Pendry},
\newblock \bibinfo{title}{Negative refraction makes a perfect lens},
\newblock \bibinfo{journal}{Phys. Rev. Lett.} \bibinfo{volume}{85}
  (\bibinfo{year}{2000}) \bibinfo{pages}{3966--3969}.
  \DOIprefix\doi{https://doi.org/10.1103/PhysRevLett.85.3966}.
%Type = Article
\bibitem[{Notomi(2002)}]{Notomi2002}
\bibinfo{author}{M.~Notomi},
\newblock \bibinfo{title}{Negative refraction in photonic crystals},
\newblock \bibinfo{journal}{Opt. Quantum Electron.} \bibinfo{volume}{34}
  (\bibinfo{year}{2002}) \bibinfo{pages}{133--143}.
  \DOIprefix\doi{https://doi.org/10.1023/A:1013300825612}.
%Type = Article
\bibitem[{Cubukcu et~al.(2003)Cubukcu, Aydin, Ozbay, Foteinopoulou, and
  Soukoulis}]{Cubukcu2003}
\bibinfo{author}{E.~Cubukcu}, \bibinfo{author}{K.~Aydin},
  \bibinfo{author}{E.~Ozbay}, \bibinfo{author}{S.~Foteinopoulou},
  \bibinfo{author}{C.~M. Soukoulis},
\newblock \bibinfo{title}{Negative refraction by photonic crystals},
\newblock \bibinfo{journal}{Nature} \bibinfo{volume}{423}
  (\bibinfo{year}{2003}) \bibinfo{pages}{604--605}.
  \DOIprefix\doi{https://doi.org/10.1038/423604b}.
%Type = Article
\bibitem[{Valentine et~al.(2008)Valentine, Zhang, Zentgraf, Ulin-Avila, Genov,
  Bartal, and Zhang}]{Valentine2008}
\bibinfo{author}{J.~Valentine}, \bibinfo{author}{S.~Zhang},
  \bibinfo{author}{T.~Zentgraf}, \bibinfo{author}{E.~Ulin-Avila},
  \bibinfo{author}{D.~A. Genov}, \bibinfo{author}{G.~Bartal},
  \bibinfo{author}{X.~Zhang},
\newblock \bibinfo{title}{Three-dimensional optical metamaterial with a
  negative refractive index},
\newblock \bibinfo{journal}{Nature} \bibinfo{volume}{455}
  (\bibinfo{year}{2008}) \bibinfo{pages}{376--379}.
  \DOIprefix\doi{https://doi.org/10.1038/nature07247}.
%Type = Article
\bibitem[{Yao et~al.(2008)Yao, Liu, Liu, Wang, Sun, Bartal, Stacy, and
  Zhang}]{Yao2008}
\bibinfo{author}{J.~Yao}, \bibinfo{author}{Z.~Liu}, \bibinfo{author}{Y.~Liu},
  \bibinfo{author}{Y.~Wang}, \bibinfo{author}{C.~Sun},
  \bibinfo{author}{G.~Bartal}, \bibinfo{author}{A.~M. Stacy},
  \bibinfo{author}{X.~Zhang},
\newblock \bibinfo{title}{Optical negative refraction in bulk metamaterials of
  nanowires},
\newblock \bibinfo{journal}{Science} \bibinfo{volume}{321}
  (\bibinfo{year}{2008}) \bibinfo{pages}{930--930}.
  \DOIprefix\doi{10.1126/science.1157566}.
%Type = Inproceedings
\bibitem[{{Iyer} and {Eleftheriades}(2002)}]{Iyer2002}
\bibinfo{author}{A.~K. {Iyer}}, \bibinfo{author}{G.~V. {Eleftheriades}},
\newblock \bibinfo{title}{Negative refractive index metamaterials supporting
  2-d waves},
\newblock in: \bibinfo{booktitle}{2002 IEEE MTTS Int. Microw. Symp. (Cat.
  No.02CH37278)}, volume~\bibinfo{volume}{2}, \bibinfo{year}{2002}, pp.
  \bibinfo{pages}{1067--1070 vol.2}.
  \DOIprefix\doi{https://doi.org/10.1109/MWSYM.2002.1011823}.
%Type = Article
\bibitem[{Smith et~al.(2004)Smith, Pendry, and Wiltshire}]{Smith2004}
\bibinfo{author}{D.~R. Smith}, \bibinfo{author}{J.~B. Pendry},
  \bibinfo{author}{M.~C.~K. Wiltshire},
\newblock \bibinfo{title}{Metamaterials and negative refractive index},
\newblock \bibinfo{journal}{Science} \bibinfo{volume}{305}
  (\bibinfo{year}{2004}) \bibinfo{pages}{788--792}.
  \DOIprefix\doi{https://doi.org/10.1126/science.1096796}.
%Type = Article
\bibitem[{Joannopoulos et~al.(1997)Joannopoulos, Villeneuve, and
  Fan}]{Joannopoulos1997}
\bibinfo{author}{J.~D. Joannopoulos}, \bibinfo{author}{P.~R. Villeneuve},
  \bibinfo{author}{S.~Fan},
\newblock \bibinfo{title}{Photonic crystals: putting a new twist on light},
\newblock \bibinfo{journal}{Nature} \bibinfo{volume}{386}
  (\bibinfo{year}{1997}) \bibinfo{pages}{143--149}.
  \DOIprefix\doi{10.1038/386143a0}.
%Type = Article
\bibitem[{Li et~al.(2007)Li, Lu, Fan, Liu, Feng, Tang, and Chen}]{Li2007}
\bibinfo{author}{J.~Li}, \bibinfo{author}{M.-H. Lu}, \bibinfo{author}{T.~Fan},
  \bibinfo{author}{X.-K. Liu}, \bibinfo{author}{L.~Feng},
  \bibinfo{author}{Y.-F. Tang}, \bibinfo{author}{Y.-F. Chen},
\newblock \bibinfo{title}{All-angle negative refraction imaging effect with
  complex two-dimensional hexagonal photonic crystals},
\newblock \bibinfo{journal}{J. Appl. Phys.} \bibinfo{volume}{102}
  (\bibinfo{year}{2007}) \bibinfo{pages}{073538}.
  \DOIprefix\doi{https://doi.org/10.1063/1.2794860}.
%Type = Article
\bibitem[{Parimi et~al.(2003)Parimi, Lu, Vodo, and Sridhar}]{Parimi2003}
\bibinfo{author}{P.~Parimi}, \bibinfo{author}{W.~Lu},
  \bibinfo{author}{P.~Vodo}, \bibinfo{author}{S.~Sridhar},
\newblock \bibinfo{title}{Photonic crystals: Imaging by flat lens using
  negative refraction},
\newblock \bibinfo{journal}{Nature} \bibinfo{volume}{426}
  (\bibinfo{year}{2003}) \bibinfo{pages}{404}.
  \DOIprefix\doi{https://doi.org/10.1038/426404a}.
%Type = Article
\bibitem[{Luo et~al.(2002)Luo, Johnson, and Joannopoulos}]{Luo2002}
\bibinfo{author}{C.~Luo}, \bibinfo{author}{S.~G. Johnson},
  \bibinfo{author}{J.~D. Joannopoulos},
\newblock \bibinfo{title}{All-angle negative refraction in a
  three-dimensionally periodic photonic crystal},
\newblock \bibinfo{journal}{Appl. Phys. Lett.} \bibinfo{volume}{81}
  (\bibinfo{year}{2002}) \bibinfo{pages}{2352--2354}.
  \DOIprefix\doi{https://doi.org/10.1063/1.1508807}.
%Type = Article
\bibitem[{Luo et~al.(2003)Luo, Johnson, Joannopoulos, and Pendry}]{Luo2003}
\bibinfo{author}{C.~Luo}, \bibinfo{author}{S.~G. Johnson},
  \bibinfo{author}{J.~D. Joannopoulos}, \bibinfo{author}{J.~B. Pendry},
\newblock \bibinfo{title}{Subwavelength imaging in photonic crystals},
\newblock \bibinfo{journal}{Phys. Rev. B} \bibinfo{volume}{68}
  (\bibinfo{year}{2003}) \bibinfo{pages}{045115}.
  \DOIprefix\doi{https://doi.org/10.1103/PhysRevB.68.045115}.
%Type = Article
\bibitem[{Kushwaha et~al.(1993)Kushwaha, Halevi, Dobrzynski, and
  Djafari-Rouhani}]{Kushwaha1993}
\bibinfo{author}{M.~S. Kushwaha}, \bibinfo{author}{P.~Halevi},
  \bibinfo{author}{L.~Dobrzynski}, \bibinfo{author}{B.~Djafari-Rouhani},
\newblock \bibinfo{title}{Acoustic band structure of periodic elastic
  composites},
\newblock \bibinfo{journal}{Phys. Rev. Lett.} \bibinfo{volume}{71}
  (\bibinfo{year}{1993}) \bibinfo{pages}{2022--2025}.
  \DOIprefix\doi{https://doi.org/10.1103/PhysRevLett.71.2022}.
%Type = Article
\bibitem[{Pennec et~al.(2010)Pennec, Vasseur, Djafari-Rouhani, Dobrzyński, and
  Deymier}]{Pennec2010}
\bibinfo{author}{Y.~Pennec}, \bibinfo{author}{J.~O. Vasseur},
  \bibinfo{author}{B.~Djafari-Rouhani}, \bibinfo{author}{L.~Dobrzyński},
  \bibinfo{author}{P.~A. Deymier},
\newblock \bibinfo{title}{Two-dimensional phononic crystals: Examples and
  applications},
\newblock \bibinfo{journal}{Surf. Sci. Rep.} \bibinfo{volume}{65}
  (\bibinfo{year}{2010}) \bibinfo{pages}{229 -- 291}.
  \DOIprefix\doi{https://doi.org/10.1016/j.surfrep.2010.08.002}.
%Type = Article
\bibitem[{Hussein et~al.(2014)Hussein, Leamy, and Ruzzene}]{Hussein2014}
\bibinfo{author}{M.~I. Hussein}, \bibinfo{author}{M.~J. Leamy},
  \bibinfo{author}{M.~Ruzzene},
\newblock \bibinfo{title}{{Dynamics of phononic materials and structures:
  historical origins, recent progress, and future outlook}},
\newblock \bibinfo{journal}{Appl. Mech. Rev.} \bibinfo{volume}{66}
  (\bibinfo{year}{2014}). \DOIprefix\doi{https://doi.org/10.1115/1.4026911},
  \bibinfo{note}{040802}.
%Type = Article
\bibitem[{Pierre et~al.(2010)Pierre, Boyko, Belliard, Vasseur, and
  Bonello}]{Pierre1997}
\bibinfo{author}{J.~Pierre}, \bibinfo{author}{O.~Boyko},
  \bibinfo{author}{L.~Belliard}, \bibinfo{author}{J.~O. Vasseur},
  \bibinfo{author}{B.~Bonello},
\newblock \bibinfo{title}{Negative refraction of zero order flexural lamb waves
  through a two-dimensional phononic crystal},
\newblock \bibinfo{journal}{Appl. Phys. Lett.} \bibinfo{volume}{97}
  (\bibinfo{year}{2010}) \bibinfo{pages}{121919}.
  \DOIprefix\doi{https://doi.org/10.1063/1.3491290}.
%Type = Article
\bibitem[{Page(2016)}]{Page2016}
\bibinfo{author}{J.~H. Page},
\newblock \bibinfo{title}{Focusing of ultrasonic waves by negative refraction
  in phononic crystals},
\newblock \bibinfo{journal}{AIP Adv.} \bibinfo{volume}{6}
  (\bibinfo{year}{2016}) \bibinfo{pages}{121606}.
  \DOIprefix\doi{https://doi.org/10.1063/1.4972204}.
%Type = Article
\bibitem[{Tol et~al.(2017)Tol, Degertekin, and Erturk}]{Tol2017PC}
\bibinfo{author}{S.~Tol}, \bibinfo{author}{F.~L. Degertekin},
  \bibinfo{author}{A.~Erturk},
\newblock \bibinfo{title}{Phononic crystal luneburg lens for omnidirectional
  elastic wave focusing and energy harvesting},
\newblock \bibinfo{journal}{Appl. Phys. Lett.} \bibinfo{volume}{111}
  (\bibinfo{year}{2017}) \bibinfo{pages}{013503}.
  \DOIprefix\doi{https://doi.org/10.1063/1.4991684}.
%Type = Article
\bibitem[{Tol et~al.(2019)Tol, Degertekin, and Erturk}]{Tol2019}
\bibinfo{author}{S.~Tol}, \bibinfo{author}{F.~Degertekin},
  \bibinfo{author}{A.~Erturk},
\newblock \bibinfo{title}{3d-printed phononic crystal lens for elastic wave
  focusing and energy harvesting},
\newblock \bibinfo{journal}{Addit. Manuf.} \bibinfo{volume}{29}
  (\bibinfo{year}{2019}) \bibinfo{pages}{100780}.
  \DOIprefix\doi{https://doi.org/10.1016/j.addma.2019.100780}.
%Type = Article
\bibitem[{Jin et~al.(2019)Jin, Djafari-Rouhani, and Torrent}]{Jin2019}
\bibinfo{author}{Y.~Jin}, \bibinfo{author}{B.~Djafari-Rouhani},
  \bibinfo{author}{D.~Torrent},
\newblock \bibinfo{title}{Gradient index phononic crystals and metamaterials},
\newblock \bibinfo{journal}{J. Nanophotonics} \bibinfo{volume}{8}
  (\bibinfo{year}{2019}) \bibinfo{pages}{685 -- 701}.
  \DOIprefix\doi{https://doi.org/10.1515/nanoph-2018-0227}.
%Type = Article
\bibitem[{Danawe et~al.(2020)Danawe, Okudan, Ozevin, and Tol}]{Danawe2020}
\bibinfo{author}{H.~Danawe}, \bibinfo{author}{G.~Okudan},
  \bibinfo{author}{D.~Ozevin}, \bibinfo{author}{S.~Tol},
\newblock \bibinfo{title}{Conformal gradient-index phononic crystal lens for
  ultrasonic wave focusing in pipe-like structures},
\newblock \bibinfo{journal}{Appl. Phys. Lett.} \bibinfo{volume}{117}
  (\bibinfo{year}{2020}) \bibinfo{pages}{021906}.
  \DOIprefix\doi{https://doi.org/10.1063/5.0012316}.
%Type = Article
\bibitem[{Carrara et~al.(2013)Carrara, Cacan, Toussaint, Leamy, Ruzzene, and
  Erturk}]{Carrara2013}
\bibinfo{author}{M.~Carrara}, \bibinfo{author}{M.~R. Cacan},
  \bibinfo{author}{J.~Toussaint}, \bibinfo{author}{M.~J. Leamy},
  \bibinfo{author}{M.~Ruzzene}, \bibinfo{author}{A.~Erturk},
\newblock \bibinfo{title}{Metamaterial-inspired structures and concepts for
  elastoacoustic wave energy harvesting},
\newblock \bibinfo{journal}{Smart Mater. Struct.} \bibinfo{volume}{22}
  (\bibinfo{year}{2013}) \bibinfo{pages}{065004}.
  \DOIprefix\doi{https://doi.org/10.1088/0964-1726/22/6/065004}.
%Type = Article
\bibitem[{Harne and Lynd(2016)}]{Harne2016}
\bibinfo{author}{R.~L. Harne}, \bibinfo{author}{D.~T. Lynd},
\newblock \bibinfo{title}{Origami acoustics: using principles of folding
  structural acoustics for simple and large focusing of sound energy},
\newblock \bibinfo{journal}{Smart Mater. Struct.} \bibinfo{volume}{25}
  (\bibinfo{year}{2016}) \bibinfo{pages}{085031}.
  \DOIprefix\doi{https://doi.org/10.1088/0964-1726/25/8/085031}.
%Type = Article
\bibitem[{Tol et~al.(2017)Tol, Degertekin, and Erturk}]{Tol2017Mirrors}
\bibinfo{author}{S.~Tol}, \bibinfo{author}{F.~L. Degertekin},
  \bibinfo{author}{A.~Erturk},
\newblock \bibinfo{title}{Structurally embedded reflectors and mirrors for
  elastic wave focusing and energy harvesting},
\newblock \bibinfo{journal}{J. Appl. Phys.} \bibinfo{volume}{122}
  (\bibinfo{year}{2017}) \bibinfo{pages}{164503}.
  \DOIprefix\doi{https://doi.org/10.1063/1.5008724}.
%Type = Article
\bibitem[{Zhang et~al.(2014)Zhang, Li, Zhu, Zhu, Yang, Wang, Yin, and
  Zhang}]{Zhang2014}
\bibinfo{author}{P.~Zhang}, \bibinfo{author}{T.~Li}, \bibinfo{author}{J.~Zhu},
  \bibinfo{author}{X.~Zhu}, \bibinfo{author}{S.~Yang},
  \bibinfo{author}{Y.~Wang}, \bibinfo{author}{X.~Yin},
  \bibinfo{author}{X.~Zhang},
\newblock \bibinfo{title}{Generation of acoustic self-bending and bottle beams
  by phase engineering},
\newblock \bibinfo{journal}{Nat. Commun.} \bibinfo{volume}{5}
  (\bibinfo{year}{2014}) \bibinfo{pages}{4316}.
  \DOIprefix\doi{https://doi.org/10.1038/ncomms5316}.
%Type = Article
\bibitem[{Zhu and Semperlotti(2016)}]{Zhu2016}
\bibinfo{author}{H.~Zhu}, \bibinfo{author}{F.~Semperlotti},
\newblock \bibinfo{title}{Anomalous refraction of acoustic guided waves in
  solids with geometrically tapered metasurfaces},
\newblock \bibinfo{journal}{Phys. Rev. Lett.} \bibinfo{volume}{117}
  (\bibinfo{year}{2016}).
  \DOIprefix\doi{https://doi.org/10.1103/PhysRevLett.117.034302}.
%Type = Article
\bibitem[{Tol et~al.(2017)Tol, Xia, Ruzzene, and Erturk}]{Tol2017SB}
\bibinfo{author}{S.~Tol}, \bibinfo{author}{Y.~Xia},
  \bibinfo{author}{M.~Ruzzene}, \bibinfo{author}{A.~Erturk},
\newblock \bibinfo{title}{Self-bending elastic waves and obstacle circumventing
  in wireless power transfer},
\newblock \bibinfo{journal}{Appl. Phys. Lett.} \bibinfo{volume}{110}
  (\bibinfo{year}{2017}) \bibinfo{pages}{163505}.
  \DOIprefix\doi{https://doi.org/10.1063/1.4981251}.
%Type = Article
\bibitem[{Darabi et~al.(2018)Darabi, Zareei, Alam, and Leamy}]{Darabi2018}
\bibinfo{author}{A.~Darabi}, \bibinfo{author}{A.~Zareei},
  \bibinfo{author}{M.-R. Alam}, \bibinfo{author}{M.~Leamy},
\newblock \bibinfo{title}{Broadband bending of flexural waves: acoustic shapes
  and patterns},
\newblock \bibinfo{journal}{Sci. Rep.} \bibinfo{volume}{8}
  (\bibinfo{year}{2018}).
  \DOIprefix\doi{https://doi.org/10.1038/s41598-018-29192-1}.
%Type = Article
\bibitem[{Su et~al.(2018)Su, Lu, and Norris}]{Su2018}
\bibinfo{author}{X.~Su}, \bibinfo{author}{Z.~Lu}, \bibinfo{author}{A.~N.
  Norris},
\newblock \bibinfo{title}{Elastic metasurfaces for splitting sv- and p-waves in
  elastic solids},
\newblock \bibinfo{journal}{J. Appl. Phys.} \bibinfo{volume}{123}
  (\bibinfo{year}{2018}) \bibinfo{pages}{091701}.
  \DOIprefix\doi{https://doi.org/10.1063/1.5007731}.
%Type = Article
\bibitem[{Sukhovich et~al.(2008)Sukhovich, Jing, and Page}]{Sukovich2008}
\bibinfo{author}{A.~Sukhovich}, \bibinfo{author}{L.~Jing},
  \bibinfo{author}{J.~H. Page},
\newblock \bibinfo{title}{Negative refraction and focusing of ultrasound in
  two-dimensional phononic crystals},
\newblock \bibinfo{journal}{Phys. Rev. B} \bibinfo{volume}{77}
  (\bibinfo{year}{2008}) \bibinfo{pages}{014301}.
  \DOIprefix\doi{https://doi.org/10.1103/PhysRevB.77.014301}.
%Type = Article
\bibitem[{Ke et~al.(2005)Ke, Liu, Qiu, Wang, Shi, Wen, and Sheng}]{Ke2005}
\bibinfo{author}{M.~Ke}, \bibinfo{author}{Z.~Liu}, \bibinfo{author}{C.~Qiu},
  \bibinfo{author}{W.~Wang}, \bibinfo{author}{J.~Shi},
  \bibinfo{author}{W.~Wen}, \bibinfo{author}{P.~Sheng},
\newblock \bibinfo{title}{Negative-refraction imaging with two-dimensional
  phononic crystals},
\newblock \bibinfo{journal}{Phys. Rev. B} \bibinfo{volume}{72}
  (\bibinfo{year}{2005}) \bibinfo{pages}{064306}.
  \DOIprefix\doi{https://doi.org/10.1103/PhysRevB.72.064306}.
%Type = Article
\bibitem[{Li et~al.(2006)Li, Liu, and Qiu}]{Li2006}
\bibinfo{author}{J.~Li}, \bibinfo{author}{Z.~Liu}, \bibinfo{author}{C.~Qiu},
\newblock \bibinfo{title}{Negative refraction imaging of acoustic waves by a
  two-dimensional three-component phononic crystal},
\newblock \bibinfo{journal}{Phys. Rev. B} \bibinfo{volume}{73}
  (\bibinfo{year}{2006}) \bibinfo{pages}{054302}.
  \DOIprefix\doi{https://doi.org/10.1103/PhysRevB.73.054302}.
%Type = Article
\bibitem[{Lu et~al.(2007)Lu, Zhang, Feng, Zhao, Chen, Mao, Zi, Zhu, Zhu, and
  Ming}]{Lu2007}
\bibinfo{author}{M.-H. Lu}, \bibinfo{author}{C.~Zhang},
  \bibinfo{author}{L.~Feng}, \bibinfo{author}{J.~Zhao}, \bibinfo{author}{Y.-F.
  Chen}, \bibinfo{author}{Y.-W. Mao}, \bibinfo{author}{J.~Zi},
  \bibinfo{author}{Y.-Y. Zhu}, \bibinfo{author}{S.-N. Zhu},
  \bibinfo{author}{N.-B. Ming},
\newblock \bibinfo{title}{Negative birefraction of acoustic waves in a sonic
  crystal},
\newblock \bibinfo{journal}{Nat. Mater.} \bibinfo{volume}{6}
  (\bibinfo{year}{2007}) \bibinfo{pages}{744--8}.
  \DOIprefix\doi{https://doi.org/10.1038/nmat1987}.
%Type = Article
\bibitem[{Morvan et~al.(2010)Morvan, Tinel, Hladky-Hennion, Vasseur, and
  Dubus}]{Morvan2010}
\bibinfo{author}{B.~Morvan}, \bibinfo{author}{A.~Tinel}, \bibinfo{author}{A.-C.
  Hladky-Hennion}, \bibinfo{author}{J.~Vasseur}, \bibinfo{author}{B.~Dubus},
\newblock \bibinfo{title}{Experimental demonstration of the negative refraction
  of a transverse elastic wave in a two-dimensional solid phononic crystal},
\newblock \bibinfo{journal}{Appl. Phys. Lett.} \bibinfo{volume}{96}
  (\bibinfo{year}{2010}) \bibinfo{pages}{101905}.
  \DOIprefix\doi{https://doi.org/10.1063/1.3302456}.
%Type = Article
\bibitem[{Cro\"enne et~al.(2011)Cro\"enne, Manga, Morvan, Tinel, Dubus,
  Vasseur, and Hladky-Hennion}]{Croenne2011}
\bibinfo{author}{C.~Cro\"enne}, \bibinfo{author}{E.~D. Manga},
  \bibinfo{author}{B.~Morvan}, \bibinfo{author}{A.~Tinel},
  \bibinfo{author}{B.~Dubus}, \bibinfo{author}{J.~Vasseur},
  \bibinfo{author}{A.-C. Hladky-Hennion},
\newblock \bibinfo{title}{Negative refraction of longitudinal waves in a
  two-dimensional solid-solid phononic crystal},
\newblock \bibinfo{journal}{Phys. Rev. B} \bibinfo{volume}{83}
  (\bibinfo{year}{2011}) \bibinfo{pages}{054301}.
  \DOIprefix\doi{https://doi.org/10.1103/PhysRevB.83.054301}.
%Type = Article
\bibitem[{Lee et~al.(2011)Lee, Ma, Lee, Kim, and Kim}]{Lee2011}
\bibinfo{author}{M.~Lee}, \bibinfo{author}{P.~Ma}, \bibinfo{author}{I.~K. Lee},
  \bibinfo{author}{H.~Kim}, \bibinfo{author}{Y.~Y. Kim},
\newblock \bibinfo{title}{Negative refraction experiments with guided
  shear-horizontal waves in thin phononic crystal plates},
\newblock \bibinfo{journal}{Appl. Phys. Lett.} \bibinfo{volume}{98}
  (\bibinfo{year}{2011}) \bibinfo{pages}{011909--011909}.
  \DOIprefix\doi{https://doi.org/10.1063/1.3533641}.
%Type = Article
\bibitem[{Zhu et~al.(2014)Zhu, Liu, Hu, Sun, and Huang}]{Zhu2014}
\bibinfo{author}{R.~Zhu}, \bibinfo{author}{X.~Liu}, \bibinfo{author}{G.~Hu},
  \bibinfo{author}{C.~Sun}, \bibinfo{author}{G.~Huang},
\newblock \bibinfo{title}{Negative refraction of elastic waves at the
  deep-subwavelength scale in a single-phase metamaterial},
\newblock \bibinfo{journal}{Nat. Commun.} \bibinfo{volume}{5}
  (\bibinfo{year}{2014}). \DOIprefix\doi{https://doi.org/10.1038/ncomms6510}.
%Type = Article
\bibitem[{Oh et~al.(2017)Oh, Seung, and Kim}]{Oh2017}
\bibinfo{author}{J.~H. Oh}, \bibinfo{author}{H.~M. Seung},
  \bibinfo{author}{Y.~Y. Kim},
\newblock \bibinfo{title}{Doubly negative isotropic elastic metamaterial for
  sub-wavelength focusing: Design and realization},
\newblock \bibinfo{journal}{J. Sound Vib.} \bibinfo{volume}{410}
  (\bibinfo{year}{2017}) \bibinfo{pages}{169--186}.
  \DOIprefix\doi{https://doi.org/10.1016/j.jsv.2017.08.027}.
%Type = Article
\bibitem[{Bramhavar et~al.(2011)Bramhavar, Prada, Maznev, Every, Norris, and
  Murray}]{Bramhavar2011}
\bibinfo{author}{S.~Bramhavar}, \bibinfo{author}{C.~Prada},
  \bibinfo{author}{A.~A. Maznev}, \bibinfo{author}{A.~G. Every},
  \bibinfo{author}{T.~B. Norris}, \bibinfo{author}{T.~W. Murray},
\newblock \bibinfo{title}{Negative refraction and focusing of elastic lamb
  waves at an interface},
\newblock \bibinfo{journal}{Phys. Rev. B} \bibinfo{volume}{83}
  (\bibinfo{year}{2011}) \bibinfo{pages}{014106}.
  \DOIprefix\doi{https://doi.org/10.1103/PhysRevB.83.014106}.
%Type = Article
\bibitem[{Dubois et~al.(2013)Dubois, Farhat, Bossy, Enoch, Guenneau, and
  Sebbah}]{Dubois2013}
\bibinfo{author}{M.~Dubois}, \bibinfo{author}{M.~Farhat},
  \bibinfo{author}{E.~Bossy}, \bibinfo{author}{S.~Enoch},
  \bibinfo{author}{S.~Guenneau}, \bibinfo{author}{P.~Sebbah},
\newblock \bibinfo{title}{Flat lens for pulse focusing of elastic waves in thin
  plates},
\newblock \bibinfo{journal}{Appl. Phys. Lett.} \bibinfo{volume}{103}
  (\bibinfo{year}{2013}) \bibinfo{pages}{071915}.
  \DOIprefix\doi{https://doi.org/10.1063/1.4818716}.
%Type = Article
\bibitem[{García and Fernandez(2015)}]{Garcia2015}
\bibinfo{author}{P.~García}, \bibinfo{author}{P.~Fernandez},
\newblock \bibinfo{title}{Floquet-bloch theory and its application to the
  dispersion curves of nonperiodic layered systems},
\newblock \bibinfo{journal}{Math. Prob. Eng.} \bibinfo{volume}{2015}
  (\bibinfo{year}{2015}). \DOIprefix\doi{https://doi.org/10.1155/2015/475364}.
%Type = Book
\bibitem[{Poruchikov(1993)}]{Poruchikov1993}
\bibinfo{author}{V.~Poruchikov}, \bibinfo{title}{Methods of the Classical
  Theory of Elastodynamics}, \bibinfo{edition}{1st} ed.,
  \bibinfo{publisher}{Springer-Verlag Berlin Heidelberg}, \bibinfo{year}{1993}.
  \DOIprefix\doi{https://doi.org/10.1007/978-3-642-77099-9}.

\end{thebibliography}






% \end{thebibliography}

\end{document}